\documentclass{ifacconf}
\usepackage{mathtools}
\usepackage{graphicx}      
\usepackage{natbib}        
\usepackage{algorithm}
\usepackage{algpseudocode}
\usepackage{multirow}
\usepackage{amsmath}
\usepackage{svg}
\usepackage{pdfpages}
\usepackage{subfig}
\usepackage{gensymb}
\usepackage{lipsum}

\begin{document}
\begin{frontmatter}

\title{Combined film and pulse heating of lithium ion batteries to improve performance in low ambient temperature}



\author[First]{Habtamu Hailemichael} 
\author[First]{Beshah Ayalew}

\address[First]{Automotive Engineering, Clemson University, Greenville, SC 29607, USA (hhailem, beshah)@clemson.edu.}

\begin{abstract}                
Low ambient temperatures significantly reduce Lithium ion batteries' (LIBs') charge/discharge power and energy capacity, and cause rapid degradation through lithium plating. These limitations can be addressed by preheating the LIB with an external heat source or by exploiting the internal heat generation through the LIB's internal impedance. Fast external heating generates large temperature gradients across the LIB due to the low thermal conductivity of the cell, while internal impedance heating (usually through AC or pulse charge/discharging) tends to be relatively slow, although it can achieve more uniform temperature distribution. This paper investigates the potential of combining externally sourced resistive film heating with bidirectional pulse heating to achieve fast preheating without causing steep temperature gradients. The LIB is modeled with the Doyle Fuller Newman (DFN) electrochemical model and 1D thermal model, and reinforcement learning (RL) is used to optimize the pulse current amplitude and film voltage concurrently. The results indicate that the optimal policy for maximizing the rate of temperature rise while limiting temperature gradients has the film heating dominate the initial phases and create the ideal conditions for pulse heating to take over. In addition, the pulse component shares the heating load and reduces the energy rating of the auxiliary power source.

\end{abstract}

\begin{keyword}
Lithium-ion battery, low temperature performance, preheating/warm up, pulse heating. 
\end{keyword}

\end{frontmatter}
\section{Introduction}
Lithium-ion batteries (LIBs) are widely adopted as energy storage devices for electric and hybrid vehicles for their rare combination of high power and energy density, high efficiency, and long cycle life. However, these benefits hold at relatively high temperatures ($>10 \degree C$), and the LIBs’ performance substantially degrades in colder temperatures. In colder temperatures, low electrolyte \citep{Smart1999} and solid-electrolyte interface (SEI) conductivity \citep{Ratnakumar2001}, sluggish solid-state diffusion \citep{Huang2000}, high charge-transfer resistance at electrolyte-electrode interface \citep{Huang2000} cause reduced power and energy capacity(in charge/discharge). Charging at low temperatures also leads to lithium plating \citep{Yang2017}, a major factor in reducing the usable life of LIB cells and may lead to lithium dendrite formation that presents safety risks (short circuits and thermal runaway) \citep{Gao2020}.

In addition to efforts to solve this problem through material and battery structure research, the main mitigating solution has been implementing temperature modulation by preheating the battery. Preheating methods can be classified as internal and external based on the heat source. Internal-heating methods take advantage of the fact that LIBs have high charge transfer impedance at low temperatures to generate ohmic heating. Internal heating by discharging the LIB is limited by the low available energy in extremely low subzero temperatures, and it is only applicable at high initial SOC conditions. Even though internal heating through charging resolves the problem of low energy availability, high C rate charging (that would also offer faster heat up) in low temperatures increases polarization voltage, making the battery degrade through lithium plating \citep{Yang2017, Yang2018}.

Bidirectional pulse heating resolves the issue of lithium plating by shuttling energy between the battery and another energy reservoir \citep{Mohan2016}. The bidirectional pulse counteracts the development of charging polarization voltage with an opposite discharge polarization \citep{Qin2020}. As it mostly shuttles the same energy, it could be used in low SOC conditions. Even if pulse heating enables heating up the battery with relatively low energy consumption and lower risk of degradation, achieving quick temperature rise is dependent, among other things, on the pulse amplitude. The speed of heating increases with the pulse amplitude, and at extremely low temperatures, achieving high C rate charging is limited by the cutoff voltages. Periodical charging and discharging with AC input is also used for AC heating which, despite providing efficient and reliable heating, is limited by the need to access an external AC source \citep{Stuart2004}.

External heating methods, on the other hand, transport heat generated externally to the LIB through convective heating using air or liquid medium \citep{Ji2013}, or conduction from resistive films placed in direct contact with the LIB \citep{LEI2015, Jin2016}. Film heating methods are attractive for their simple installation and heating efficiency. However, providing fast and uniform heating requires several films placed strategically in a module (pack), considerably increasing volume and mass. On the other hand, using fewer films results in localized heating, creating a significant temperature gradient between the portion of the LIBs with direct contact with the film and the core. Heating films considered in this paper are powered by an auxiliary power source, which could be a well-maintained smaller battery or a supercapacitor with a sufficient energy rating. Such supercapacitors are becoming available in the market utilizing hybrid technology and graphene plates \citep{Muzaffar2019}.

In this paper, we investigate combining bidirectional pulse heating with film heating to mitigate the individual shortcomings of either method and exploit their strengths. Film heating can quickly bring the battery to states where we can use a high-amplitude pulse. On the other hand, for the same temperature rise rate (RTR), combining external film and pulse heating can reduce the temperature gradient compared to the case where equivalent energy is supplied by only the external film. Since they share the heating load, the auxiliary power source's energy rating can be lower than what would have been for only film heating. We assume that the auxiliary power source is well-insulated and maintained near a full charge, and as a result, such a system could still be used in low SOC conditions of the main LIB pack. In addition to providing energy for film heating, the auxiliary power source could also be used as a reservoir to shuttle energy in pulse heating. 

The proposed preheating system consists of bidirectional pulse heating combined with external heating via a positive temperature coefficient resistive film (PTC) that modulates its resistance to prevent thermal overheating \citep{Jin2016}. While film heating component is controlled by the PTC voltage, achieving effective pulse heating and lithium plating minimization requires pulse amplitude, frequency, and duty cycle modulation. Optimization of these combined heating mechanisms should consider the nonlinearly coupled electrochemical and thermal dynamics which are not fully observable. Model-based preheating optimizations simplify the electrochemical and thermal models to a manageable scale with reduced electrochemical or equivalent circuit models \citep{Du2018, Mohan2016}. These simplifications make optimization manageable, but results become inaccurate in the case of modeling error and uncertainty. In recent times, Reinforcement learning (RL) has been proposed to capture the complexity of LIBs and introduce customizability to control policies as the LIB parameters change with age. For example, RL has recently been applied for extracting optimal fast charging policies \citep{Park2020}. In this paper, we formulate the heating process as a Markov Decision Process (MDP) to which we apply an RL approach for joint optimization for the two heating methods and gain insight into the resulting optimal heating policies for achieving fast and uniform preheating of LIBs.

The rest of the paper is organized as follows. Section \ref{sec: II} describes the electrochemical and thermal modeling of the LIB. Section \ref{sec: III} describes the architecture of an RL agent used to optimize the PTC voltage and pulse current amplitude. Section \ref{sec: IV} discusses the implementation RL optimization on a simulated LIB cell, and Section \ref{sec: V} concludes the paper.

\section{Electrochemical and thermal modeling} \label{sec: II}

The LIB module considered is made of pouch cells separated by aluminum plates to which the PTC film heaters are attached. Each cell has multiple units separated by separators. The electrochemistry of a unit is modeled with the Doyle Fuller Newman (DFN) model \citep{DFN2002, Plett2015}. Table \ref{tb:Nomenclature} in the Appendix gives the nomenclature for electrochemical model states and parameters. The lithium concentration in the electrolyte ($c_e (x,t)$) and solid ($c_s(x,r,t)$) is captured through a volume-averaged approximation of mass conservation.
\begin{equation} \label{eq:DFNmasselec}
\begin{split}
{{
\frac{\varepsilon_e c_e }{\partial t}=\nabla\bullet(D_{e,eff}\ \nabla c_e\ )+a_s(1-t_+^0\ )j,
}}
\end{split}
\end{equation}
\begin{equation} \label{eq:outofrange}
\begin{split}
{{
\frac{\partial c_s}{\partial t}=\frac{1}{r^2}\frac{\partial}{\partial r}\left(D_sr^2\frac{\partial c_s}{\partial r}\right).
}}
\end{split}
\end{equation}

The volume-averaged approximation for charge conservation in the solid $(\phi_{s})$ and electrolyte $(\phi_{e})$ phase of the porous electrode gives 
\begin{equation} \label{eq:outofrange}
\begin{split}
{{
\mathrm{\nabla}.\left(\sigma_{eff}\mathrm{\nabla}\phi_s\right)=a_sF{j},
}}
\end{split}
\end{equation}
\begin{equation} \label{eq:outofrange}
\begin{split}
{{
\nabla\bullet\left(k_{eff}\nabla\phi_e+k_{D,eff}\nabla \ln{c_e}\ \right)+a_sF{j}=0.
}}
\end{split}
\end{equation}

The volume-averaged approximation to the microscopic Butler–Volmer kinetics relationship gives

\begin{equation} \label{eq:outofrange}
\begin{aligned}
& j = \frac{i_{0}}{F}\left\{ \exp\left(\frac{\left(1-\alpha\right)F}{RT}\eta\right) - \exp\left(-\frac{\alpha F}{RT}\eta\right) \right\}, \\
& i_0 = Fk_0c_e^{1-\alpha}c_{s,\text{max}}\left(\frac{c_e}{c_{e,0}}\right)^{1-\alpha}\left(\frac{c_{s,\text{max}}-c_{s,e}}{c_{s,\text{max}}}\right) \\
& \quad \times \left(\frac{c_{s,e}}{c_{s,\text{max}}}\right)^\alpha.
\end{aligned}
\end{equation}

where $D_{e,eff}$, $\sigma_{eff}$,$k_{eff}$ are the effective electrolyte diffusivity, solid conductivity and electrolyte conductivity respectively. These PDEs are bound by the boundary conditions of zero lithium diffusion at the boundary, mass conservation at the solid-electrolyte interface and the center of particle, and zero ionic current at the current collectors boundary conditions.

\begin{figure}%
    \centering
    \subfloat[\centering LIB unit \citep{Plett2015}]{{\includegraphics[width=5.5cm, height=3.0cm]{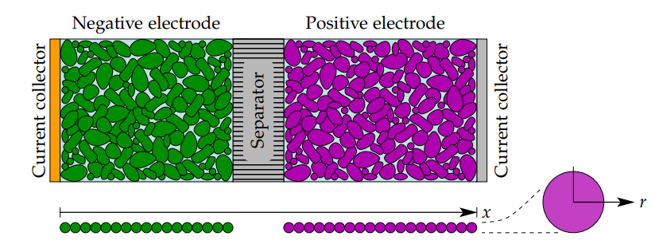} }}%
    \subfloat[\centering LIB cell]{{\includegraphics[width=4cm, height=4cm]{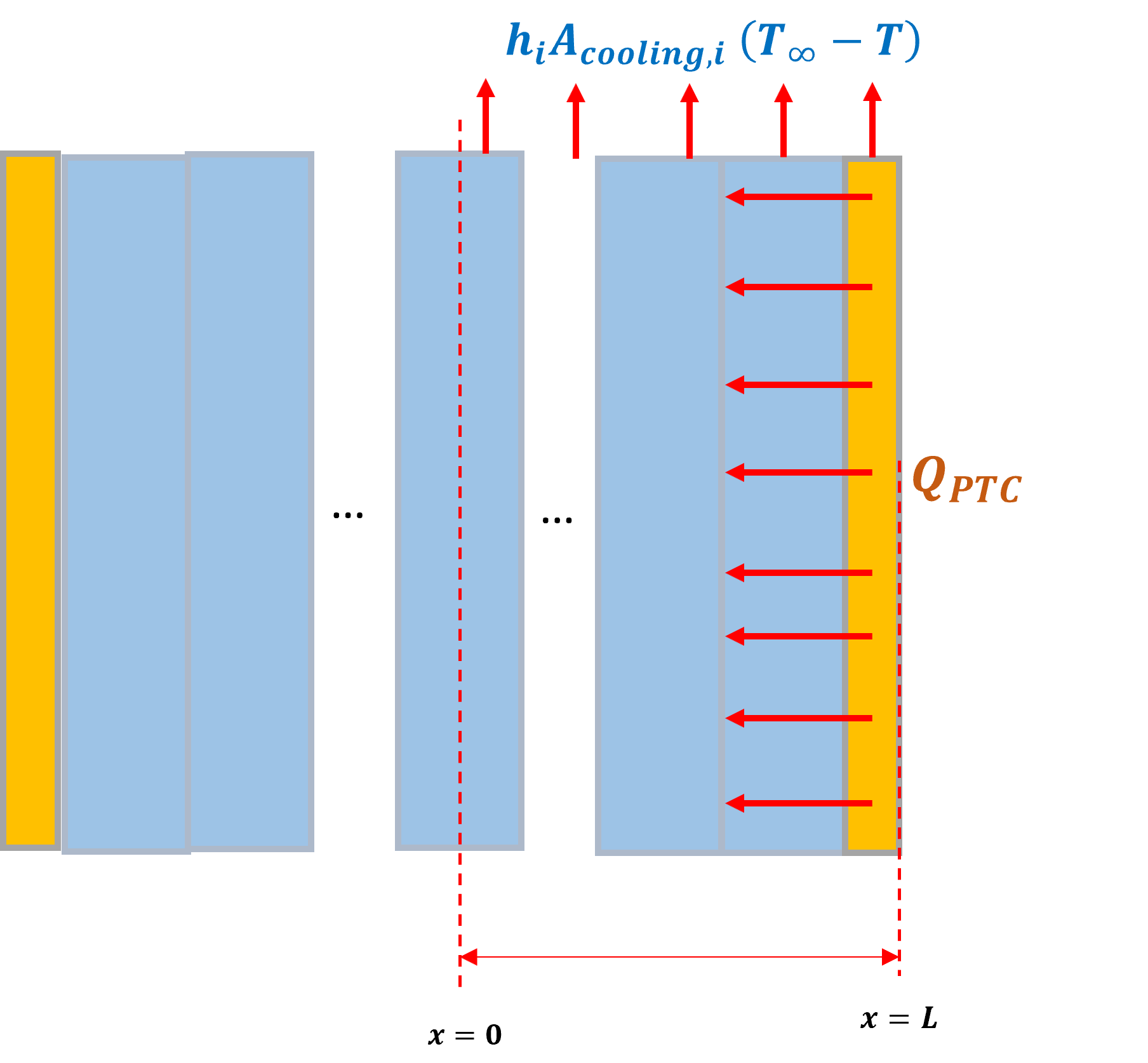} }}%
    \caption{LIB unit and cell used for electrochemical and thermal modeling}%
    \label{fig:example}%
\label{fig:celltheraml}
\end{figure}

The thermal analysis considers the temperature variation across the pouch cell (x direction) as shown in Fig.\ref{fig:celltheraml}b. On both sides of the cell, the PTC act as a thermal source. Electrochemical heat generation is also considered using a single representative unit with x-averaged temperature $(T_{avg})$ to reduce the computation burden. Heat is assumed to escape to the ambient through convection across the edge of the cell.  Considering symmetry in the x direction, the thermal model for the half cell is given as:

\begin{equation} \label{eq:outofrange}
\begin{split}
& \rho c \frac{\partial T}{\partial t} = \lambda \nabla^2 T + Q_{\text{gen}}(T_{\text{avg}}) + \frac{2\left(l_y + l_z\right)}{l_y l_z} h \left(T_\infty - T\right), \\
& T_{\text{avg}} = \frac{1}{L} \int_{0}^{L} T(x) \, dx
\end{split}
\end{equation}

with boundary conditions in the median and outer section of the cell:
\begin{equation} \label{eq:outofrange}
\begin{split}
{{
 \frac{dT}{dx}\bigg\rvert_{x=0}=0,
}}
\end{split}
\end{equation}
\begin{equation} \label{eq:outofrange}
\begin{split}
{{
 \frac{dT}{dx}\bigg\rvert_{x=L}=\frac{Q_{PTC}/2}{\lambda_{cn}A_{yz}}.
}}
\end{split}
\end{equation}

where $\lambda, h, \rho , c, A_{yz}$ are the thermal conductivity, convective heat transfer coefficient, density, specific heat capacity and cross-sectional area of the cell, respectively. The internal heat generation ($Q_{gen}$) includes reversible and irreversible generation due to chemical reactions, and joule heating due to electrical potential gradient in the solid and electrolyte.
\begin{equation} \label{eq:outofrange}
\begin{split}
Q_{gen}\left( T_{\text{avg}} \right) = & \, a_s F j \left( \eta + T_{\text{avg}} \frac{\partial U_{\text{ocp}}}{\partial T} \right) 
+ \sigma_{\text{eff}} \left( \nabla \phi_s \cdot \nabla \phi_s \right) \\
& + k_{\text{eff}} \left( \nabla \phi_e \cdot \nabla \phi_e \right) 
 + k_{D,\text{eff}} \left( \nabla \ln{c_e} \cdot \nabla \phi_e \right)
\end{split}
\end{equation}

We use the ohmic power relationship for the given voltage to model the heat generated by the PTC ($Q_{PTC}$) (\ref{eq:ohmic}). Considering aluminum plate as a perfect conductor, the temperature dependence of the resistance (and as a result, the power generated) is given by the relation:
\begin{equation} \label{eq:ohmic}
\begin{split}
{{
Q_{PTC}=\frac{v_{PTC}^2}{R_{PTC}}.\ \ \ \ 
}}
\end{split}
\end{equation}
\begin{equation} \label{eq:outofrange}
\begin{split}
R_{PTC} = 
& \begin{cases} 
      R_0 \exp\left( \alpha_0 (T - T_0) \right) & \text{for } T < T_c \\
      R_0 \exp\left( \alpha_1 (T - T_1) \right) & \text{for } T \ge T_c 
   \end{cases}
\end{split}
\end{equation}

where $R_0$ and $R_1$ are the resistance at nominal temperature $T_0$ and reference temperature $T_1$, respectively, and the Curie temperature $T_c$ is given by 
\begin{equation} \label{eq:outofrange}
\begin{split}
{{
Tc\ =\frac{\log{\left(R_1\right)}-\log{\left(R_0\right)}+\ \alpha_0T_0\ -\ \alpha_1T_1}{\alpha_0\ -\ \alpha_1}}}
\end{split}
\end{equation}

\section{Formulation of a Reinforcement Learning Agent } \label{sec: III}
\begin{figure}
\begin{center}
\includegraphics[width=8.4cm, height=5cm]{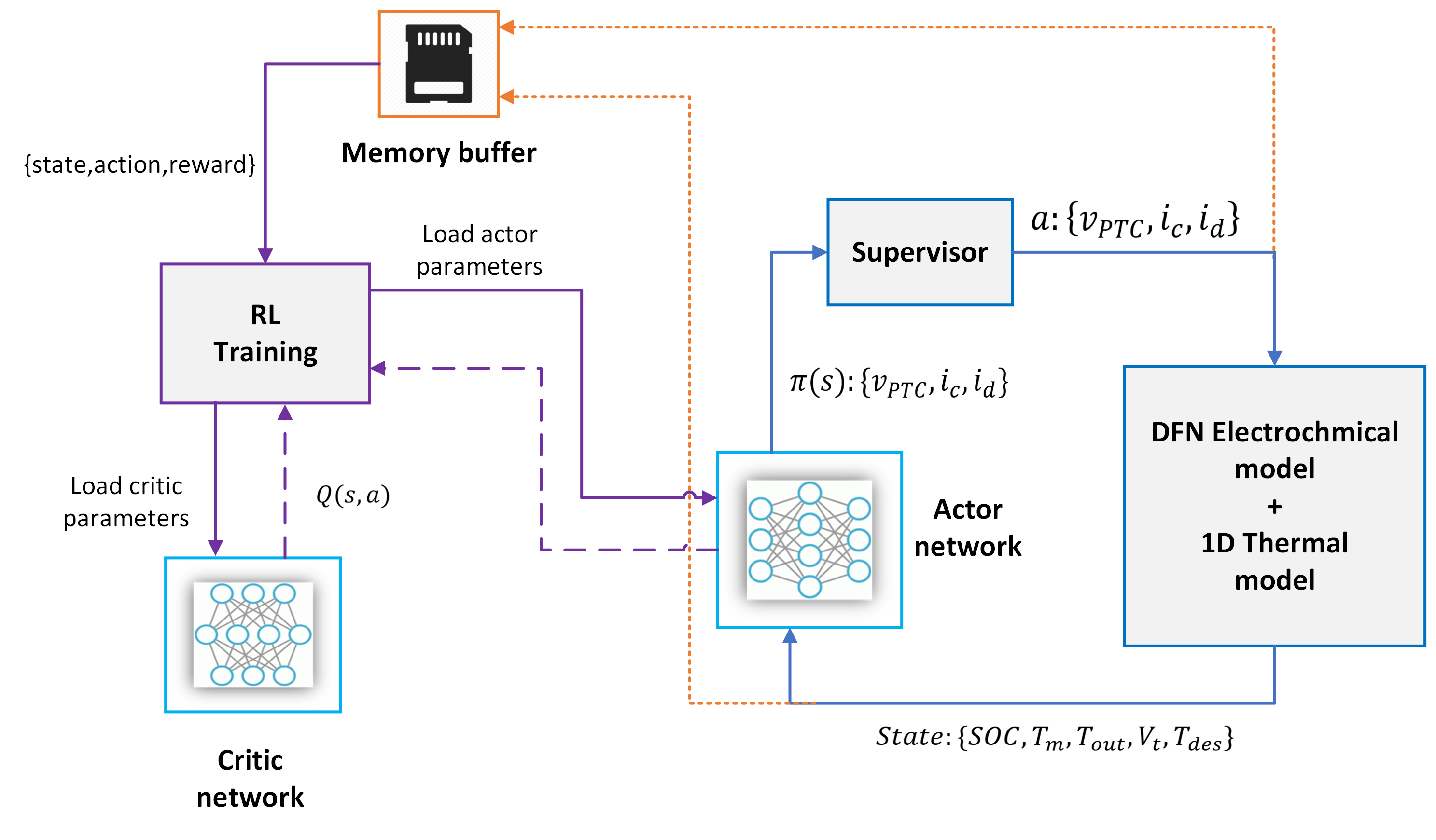}  
\caption{Actor-Critic RL for optimal preheating} 
\label{fig:RLNetwork}
\end{center}
\end{figure}
We formulate the battery preheating process as a Markov Decision Process (MDP) with state $s$, actions $a$ reward function $r(s,a)$, and discounting factor $\gamma$. The RL schematic is shown in Fig.\ref{fig:RLNetwork}. The observations of SOC, terminal voltage, the median and outer temperatures conjugated with the desired temperature constitute the state $s\ =\{SOC, T_m, T_{out}, v_t, T_{des}\}$. By fixing the pulse's duty cycle and frequency, we reduce the optimization variables and the control action ends up being PTC voltage, pulse charge and discharge amplitudes $a\ =\{v_{PTC},\ i_c,\ i_d \} $. To counteract the buildup of charging polarization (which causes lithium plating), the predetermined frequency is set to be relatively high \citep{Qin2020}, and the discharge current's magnitude is always set to be higher than the charging current in each pulse cycle \citep{Mohan2016}. Such pulse currents amplitude setting, along with the fact that the peak charging amplitude is limited by the low temperature ambient, results in a reduction in the SOC of the main LIB. 

The RL has actor-critic architecture in which the actor proposes the control action, and the critic measures how good the actions are for the given state. The actor network outputs the mean and variance values of each action, and the combination of the actions proposed by the RL $(\{v_{PTC},\ i_c,\ i_d\})$ is then sampled from multidimensional normal distribution.
\begin{equation} \label{eq:torqeguassian}
{
\pi^{T}_{\boldsymbol\theta}\left(a|s\right)=\mathcal{N}\left(\mu_{\boldsymbol\theta}\left(s\right),\sigma_{\boldsymbol\theta}^2\left(s\right)\right) 
},
\end{equation}

where $\mu, \sigma^2$ and $\theta$ are the mean, variance and parameters of the actor network, respectively. The proposed actions will then be applied to the electrochemical and thermal model of the battery environment given in Section \ref{sec: II}. However, before sending these action to the battery environment, the supervisor makes sure the actions proposed are feasible and safe. This includes ensuring the charging current is less than the maximum possible amount for the LIB's SOC and temperature, and the discharge current amplitude is higher than the charging current amplitude.

After observing the effect of the actions implemented in the simulated LIB environment, a reward is assigned for the state and action combination. We defined the reward function to promote fast heating without creating an extreme temperature gradient across the cell as measured by the difference between the outside ($T_{out}$) and core temperature ($T_m$). 

\begin{equation}\label{eq:reward}
\begin{aligned}
& R = w_R r_{RTR} + r_g, \\
& r_{RTR} = \frac{T_{\text{avg},t} - T_{\text{avg},t-1}}{\Delta t}, \\
& \text{If } T_{\text{range}} > T_t, \\
& \quad r_g = -w_{Tr} T_{\text{range}} - w_{\Delta T} \Delta T_{\text{range}}.
\end{aligned}
\end{equation}

\begin{equation} \label{eq:finalstate}
\begin{aligned}
& \text{If } T_{\text{range}} < T_t, \\
& \quad r_g = r_{\text{term}}, \\
& \text{If } T_{\text{range}} > T_t, \\
& \quad r_g = -r_{\text{term}},
\end{aligned}
\end{equation}

where $r_{RTR}$ is a reward term that encodes the average relative temperature rise (RTR) of the cell in a given time step $\Delta t$, and $r_g$ modulates the temperature gradient when it is beyond a certain threshold $T_t$. $r_g$ penalizes steep spacial temperature gradient represented by $T_{range} = T_{out} – T_{m}$ and rewards actions that reduce the temporal temperature gradient represented by $\Delta T_{range} = \Delta T_{range, t} - \Delta T_{range, t-1}$. Each reward term is then weighted by the corresponding weights $\{w_R, w_{Tr}, w_{\Delta T}\}$ to tune the overall reward according to the needs of the heating control. At the end of the heating process (the terminal state), however, $r_g$ is used to incur a big penalty $(-r_{term})$ or reward $(r_{term})$ based on the final temperature gradient (\ref{eq:finalstate}).

The state, action and reward tuple is stored in a memory/replay buffer later to be used for training. We used the Maximum A Postriori Policy optimization (MPO), an off-policy algorithm that trains the neural network in a data-efficient manner. Other state-of-the-art deep RL algorithms such as Proximal Policy Optimization (PPO) \citep{Schulman2017} are also applicable for this purpose. We refer the reader to ~\cite{Abdolmaleki2018} for implementation details of MPO.


\begin{figure*}%
    \centering
    \subfloat[\centering PTC voltage and pulse amplitude profile]{{\includegraphics[width=6.0cm, height=4.5cm]{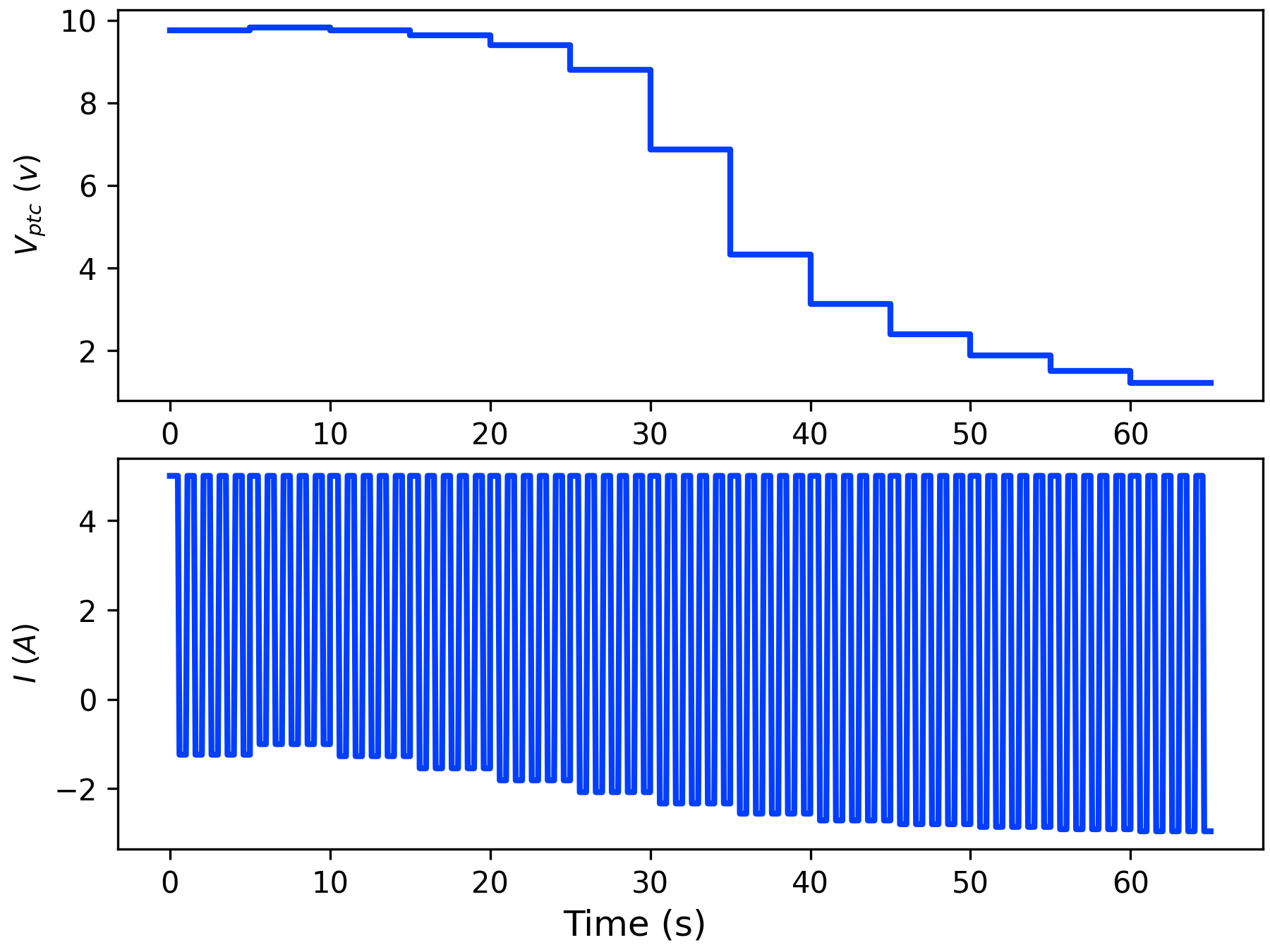} }}%
    \subfloat[\centering Heating power]{{\includegraphics[width=6.0cm, height=4.5cm]{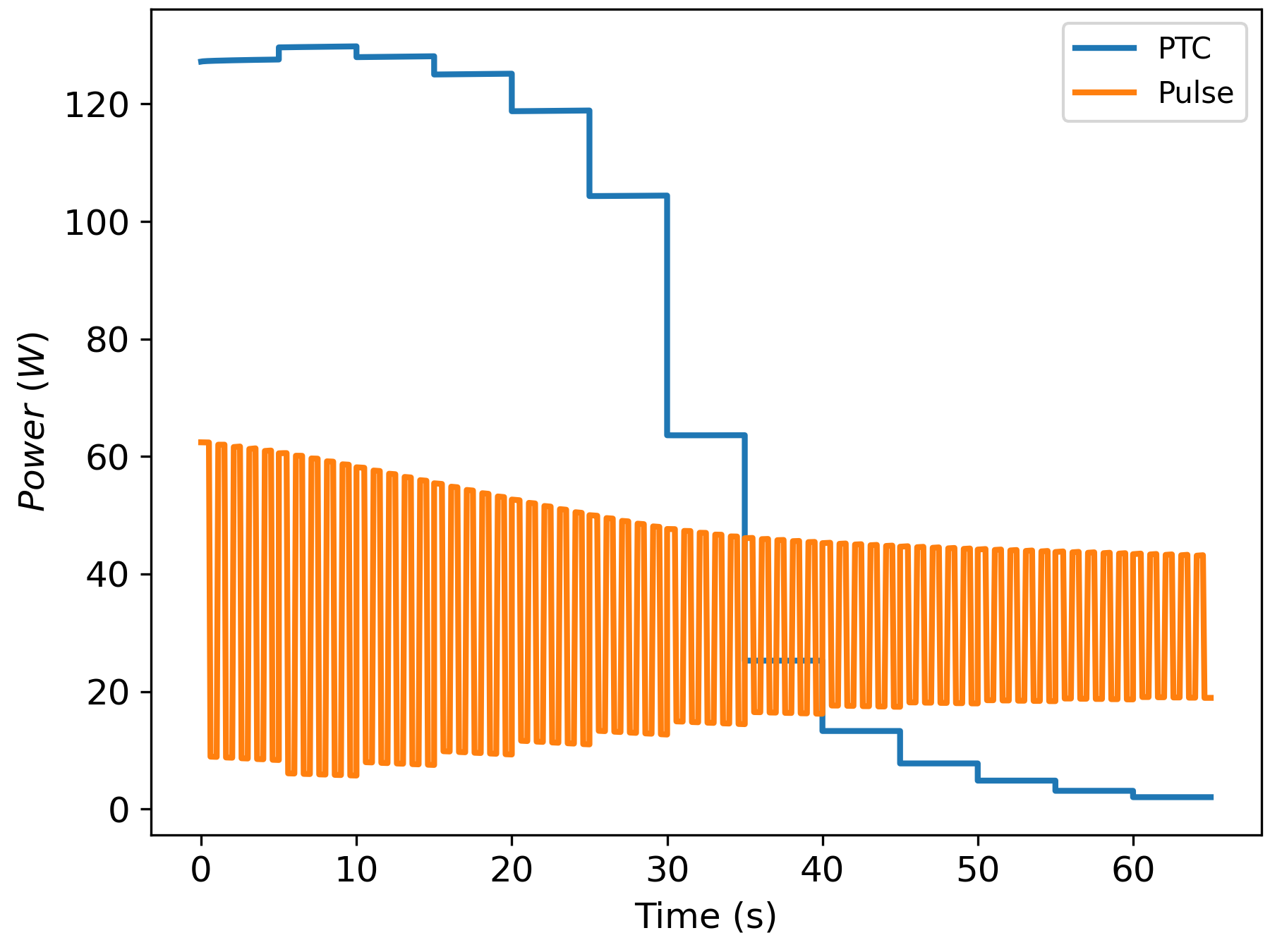} }}%
    \subfloat[\centering Temperature gradient]{{\includegraphics[width=6.0cm, height=4.5cm]{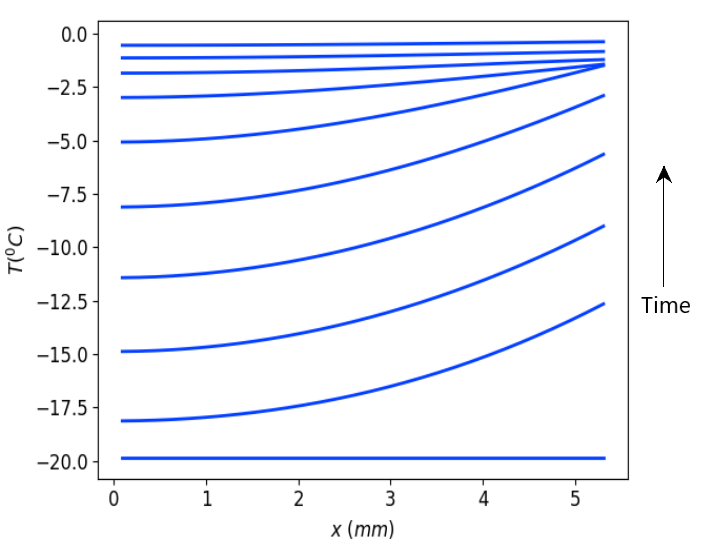} }}%
\caption{ \centering  Optimization of pulse and PTC heating (with $v_{max} = 10V$) for LIB cell heating}%
\label{fig:PTCpulse}    
\end{figure*}

\begin{figure*}%
    \centering
    \subfloat[\centering Average temperature]{{\includegraphics[width=6.0cm, height=4.5cm]{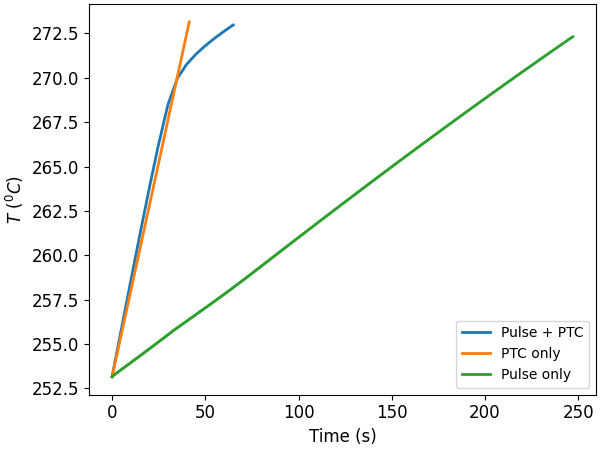} }}%
    \subfloat[\centering Cell temperature]{{\includegraphics[width=6.0cm, height=4.5cm]{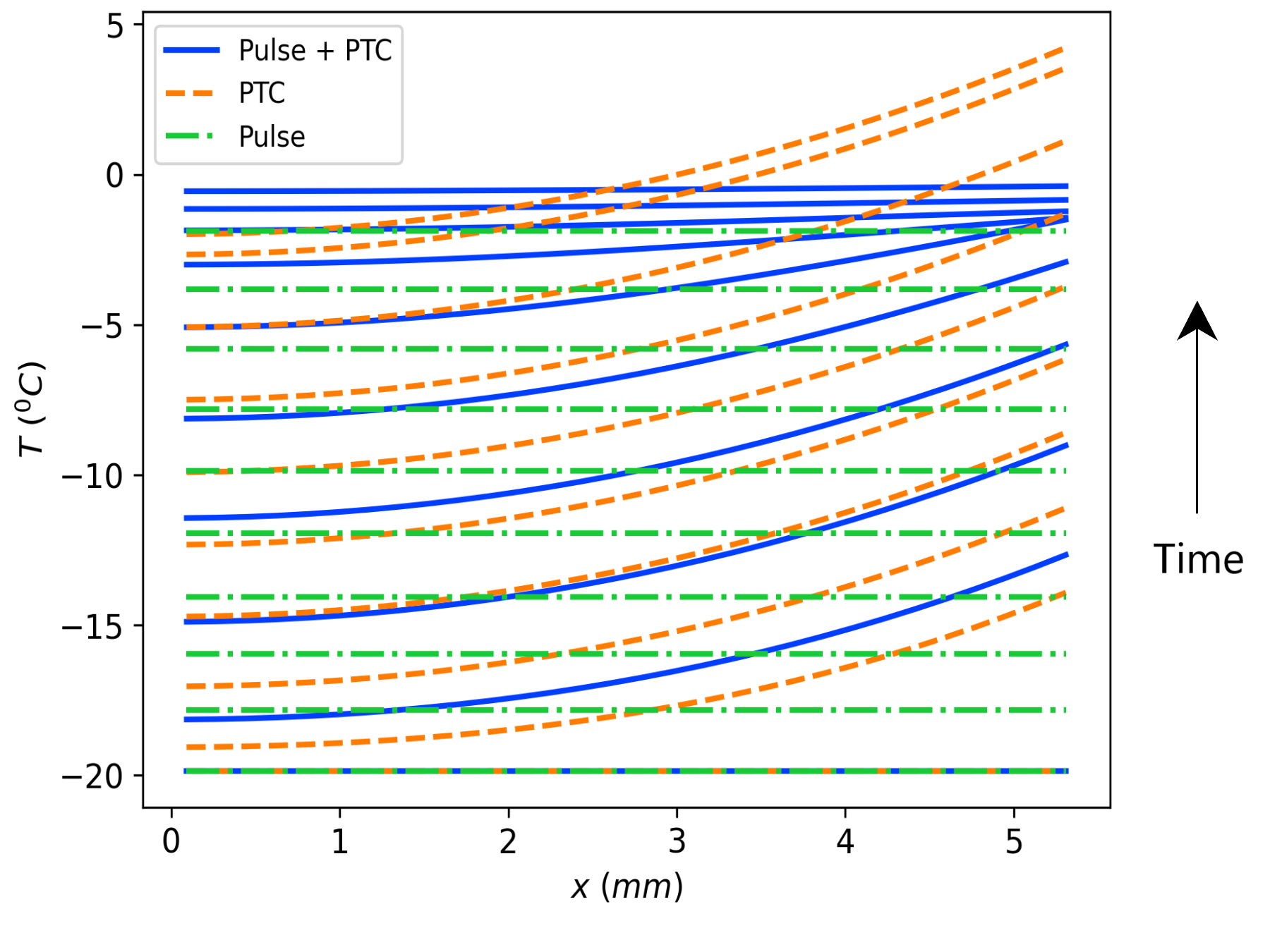} }}%
    \subfloat[\centering SOC levels]{{\includegraphics[width=6.0cm, height=4.5cm]{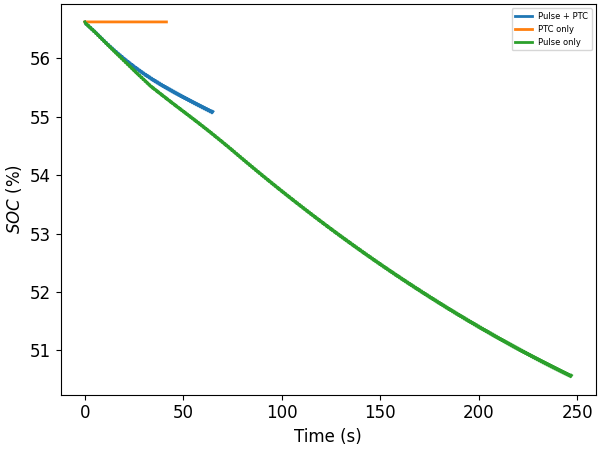} }}%
    \caption{ \centering Comparison of different preheating methods}%
\label{fig:compare} 
\end{figure*}

\section{Results and Discussions } \label{sec: IV}

The performance and characteristics of the combined PTC film and pulse heating is investigated on the LIB pouch cell model. The LIB cell has LiPF6 cell chemistry with a 0.68 Ah rating. It consists of 34 units, each with positive and negative electrodes, a separator, and positive and negative current collectors. The electrochemical properties and the geometry of the cell units are taken from \cite{Marquis2019}. The cell units alongside the aluminum plate constitute a thickness of $10.6 mm $, and the cell has $0.137 m \times 0.207 m$ cross-section. The aluminum plates are attached to PTC on each side, and the thermal parameters and the characteristic of the PTC in the experiment are given in Table \ref{tb:thermal}. The thermal model of the cell and the electrochemical model of the LIB units are simulated with Python Battery Mathematical Modelling (PyBaMM) \citep{Sulzer2021} using finite volume analysis by discretizing them into a 1D mesh.

\begin{table}[hb]
\begin{center}
\caption{PTC and thermal parameters }\label{tb:thermal}
\begin{tabular}{cc|cc}
\hline
\multicolumn{2}{| c |}{PTC Parameters} & \multicolumn{2}{| c |}{Thermal parameters of the cell} \\\hline
$R_0 \left(\Omega \right)$ & 0.367 & $h\left(W m^{-2}K^{-1}\right)$& 10 \\
$T_1 \left(K\right)$& $393$ & $ t_{Al}\left(mm\right)$ & 1 \\
$\alpha_0 \left(K^{-1}\right)$ & $-0.0044$ & $\lambda \left(W m^{-1}K^{-1}\right)$ & 2 \\
$\alpha_1 \left(K^{-1}\right) $ & $0.937$ & $\rho c\left(J m^{-3}K^{-1}\right)$ & $1.8 \times 10^6$ \\
\hline
\end{tabular}
\end{center}
\end{table}

The RL agent has feedforward actor and critic networks, each consisting of three layers of 64 nodes. It is trained to bring the battery to zero $0^oC$ x-averaged temperature from random initial temperatures down to $-20^oC$ and different SOC levels by controlling the pulse current amplitudes and the PTC voltage. A fixed pulse frequency of 1Hz and duty cycle of $50\%$ are used, and a selected pulse configuration is kept for a period of 5 seconds. The objective of achieving rapid temperature rise without causing a significant temperature gradient is encoded into the reward using a weight of $\{w_{RTR} = 1. w_{Tr} = {2}, w_{\Delta T} = 1, r_{term} = 200 \}$, and the threshold temperature difference between the core and the outer section is set to be 0.5 $(T_t = 0.5)$. States, RL actions, and the associated rewards are collected into a memory buffer and used to train the actor and critic networks by selecting random batches. 

Figure \ref{fig:PTCpulse} shows RL training results of setting the cell thermal conductivity to $2W/mK$ and maximum PTC voltage to $10V$. The training results in a heating policy in which the first phase is dominated by PTC heating. Such fast heating quickly increases the temperature and creates an ideal LIB state for the pulse heating to use high charging amplitudes safely. However, the fast temperature increase is accompanied by a high-temperature gradient. The reward is structured to lightly penalize temperature gradients initially, and as the LIB is close to the terminal state, PTC heating is reduced to avoid a high terminal temperature gradient penalty. The later phase of the heating process is dominated by pulse heating, which increases the temperature relatively uniformly. This phase also gives time to distribute the temperature gradient created by the PTC heating.

Figure \ref{fig:compare} compares the performance of combined PTC + pulse heating with PTC-only heating with the maximum voltage $(V_{PTC} = 10 V)$ and pulse heating with the maximum possible charging and discharge current. The PTC raises the average temperature the fastest but it cause a significant temperature difference between the core and the outer layer. The pulse heating takes very long as compared to the other methods, and causes a larger SOC reduction in the main LIB. Combining the pulse and PTC heatings offers compromises by achieving a relatively fast temperature rise without causing big temperature gradient. Furthermore, the energy required for the heating is shared between the LIB and the auxiliary power source and, and as shown in Table \ref{tb:energy}, reduces the energy rating of the auxiliary power source. The table also shows that pulse only case has the highest energy requirement to heat up the same temperature range due to the higher convective loss to the ambient associated with longer heating time.

\begin{figure*}%
  \centering
    \subfloat[\centering PTC component voltage profile]{{\includegraphics[width=5.5cm, height=4.5cm]{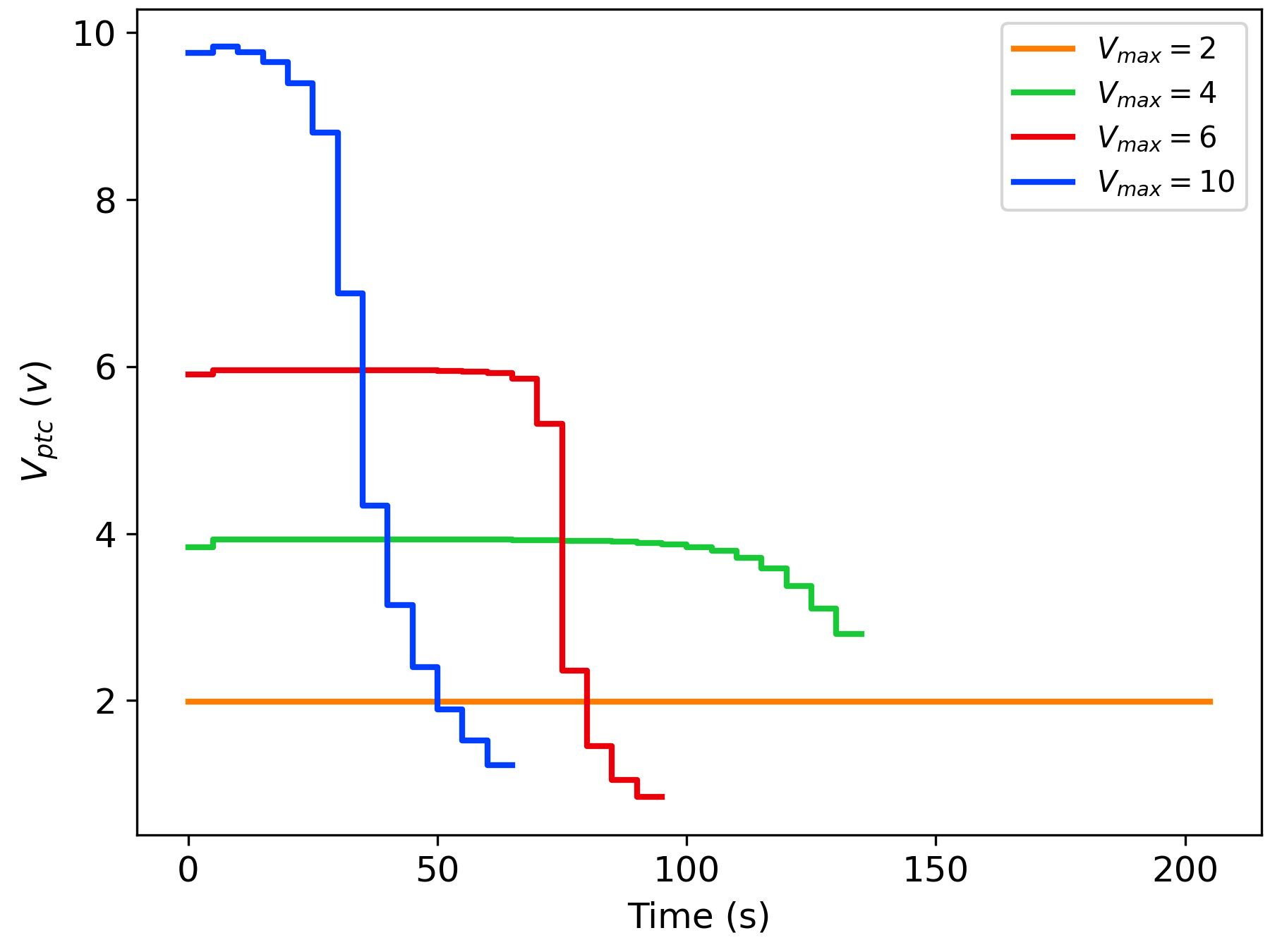} }}%
    \subfloat[\centering Pulse component current profile]{{\includegraphics[width=7.0cm, height=4.5cm]{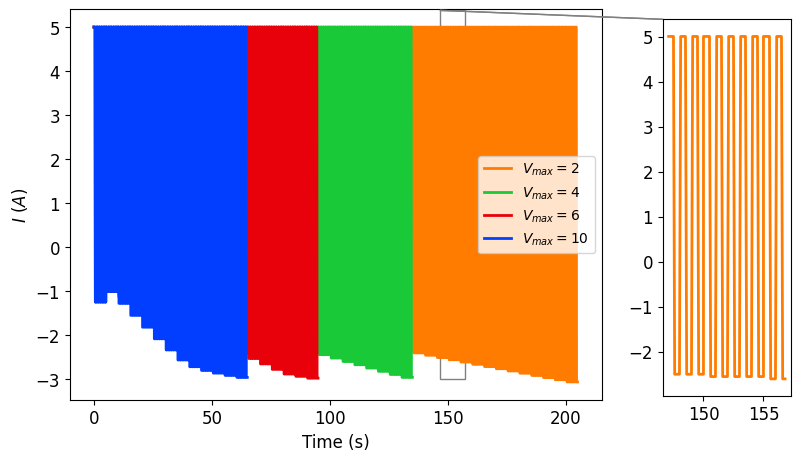} }}%
        \subfloat[\centering Average temperature]{{\includegraphics[width=5.5cm, height=4.5cm]{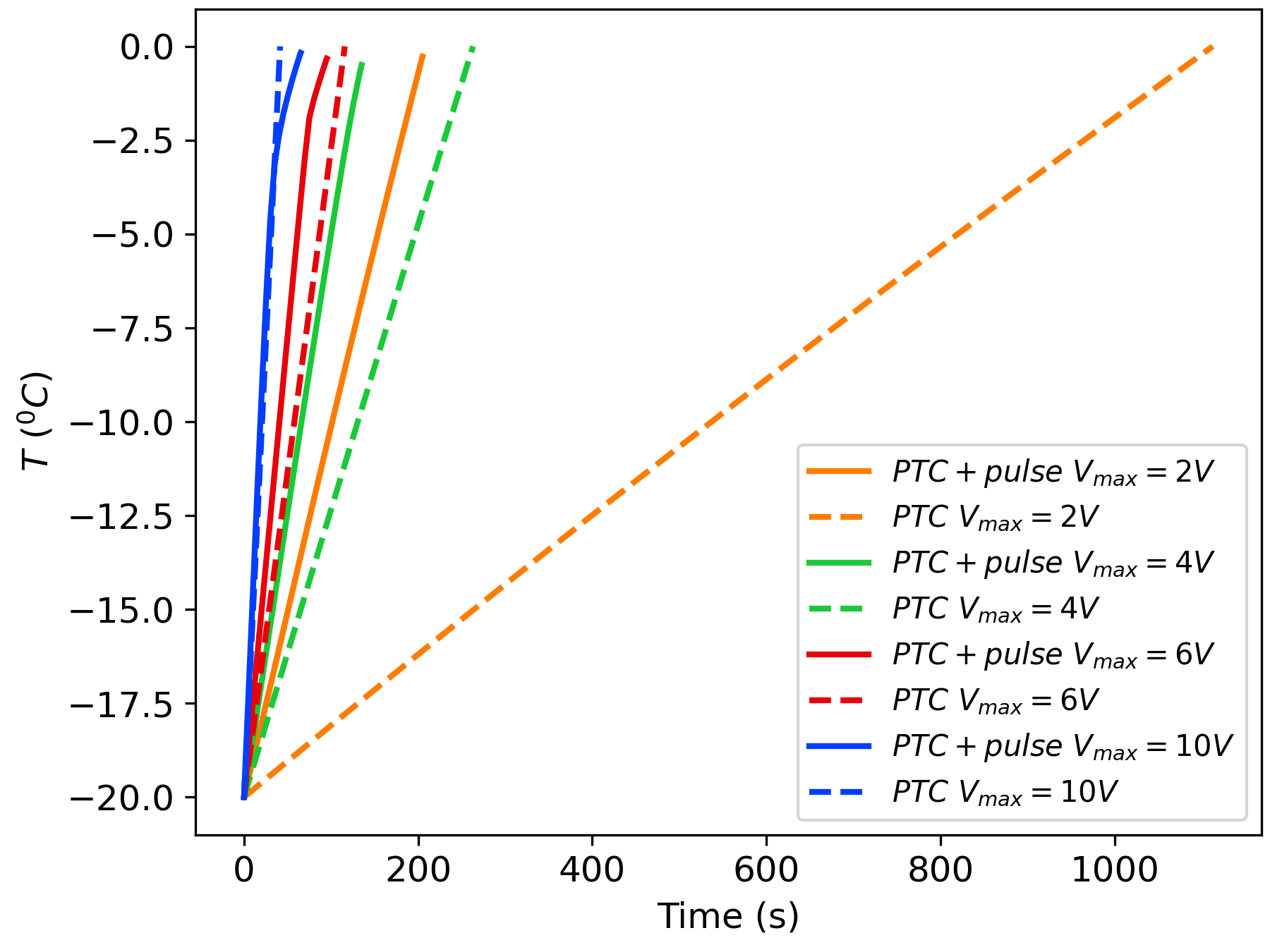} }}%
    \caption{ \centering PTC and pulse heating for different maximum PTC voltages}%
\label{fig:diffvolt} 
\end{figure*}

We also investigate what happens as the maximum voltage of the PTC is changed in Fig.\ref{fig:diffvolt}. Reducing the maximum limiting voltage of the PTC, even if it reduces the RTR, also lowers the chance of creating a big temperature gradient. As shown in Fig.\ref{fig:diffvolt}a, the optimization suggests applying the maximum voltage throughout the heating period when the maximum voltage of the PTC is set to lower values. It is also possible to notice from Fig.\ref{fig:diffvolt}c that the portion of the heating by the pulse increases as the maximum PTC voltage decreases, and after a point, the pulse and PTC combined heating has the highest RTR (as compared to the PTC only and pulse only cases). For all maximum voltage levels of the PTC, the heating optimization results in a pulse heating profile with the peak safe current amplitude (determined by the supervisor) and the PTC voltage profile that can be captured by simple rules (tables) for practical implementation.
\begin{table}[hb]
\begin{center}
\caption{Energy requirements PTC-only heating, pulse-only heating and a combined PTC and pulse heating}
\label{tb:energy}
\begin{tabular}{p{1.5cm}|p{2cm}p{1.75cm}p{1.75cm}}
\hline
& PTC component $(J)$ & Pulse \par component $(J)$ & Total \par $(J)$\\\hline
 PTC only & {\quad}$ 5563$ &{\quad} 0 & 5563\\
PTC and Pulse & {\quad} 4266 &{\quad}2182 & 6448 \\ 
Pulse only & {\quad} 0 &{\quad}9628 &  9628 \\ \hline 

\end{tabular}
\end{center}
\end{table}

\section{Conclusion} \label{sec: V}
This paper investigates the preheating of lithium-ion batteries with combined external resistive PTC film heating and internal pulse charge/discharge heating. The LIB cell, consisting of layers of pouch units and resistive heating films at each side, is modeled with DFN electrochemical model combined with a 1D thermal model. Deep reinforcement learning is used to optimize the amplitude of the pulse heating and the voltage of the PTC to achieve fast heating while avoiding a significant temperature gradients in the LIB. The optimization results demonstrate that it is possible to achieve a fast temperature rise without causing a significant temperature gradient by modulating the pulse and PTC control parameters. The two heating methods complement each other in the sense that the pulse component help to reduce the energy rating of the auxiliary power source for the PTC heating, and the PTC heating help to achieve effective pulse heating by quickly bring the LIB to states ideal for pulse heating. The RL optimization of the combined heating outputs the schedule of a maximum possible pulse current amplitude (for LIB's temperature and SOC) and a PTC voltage profile that can be translated into a simple rule for practical implementations.

Future works includes experimental verification of these finds on a prototype system as well as further pareto front studies to clarify trade-offs in the design of the system.

\begin{table}
\caption{APPENDIX: NOMENCLATURE}\label{tb:Nomenclature}
\begin{tabular}{l|ll}

\cline{1-2}
\multicolumn{2}{l}{ Electrochemical model states and parameters}  \\ 
\cline{1-2}
$c_s^\pm$ & Lithium concentration in solid phase ${[}mol/m^3{]}$ & \\
$c_e$ & Lithium concentration in electrolyte phase ${[}mol/m^3{]}$ & \\
$\phi_s^\pm$ & Solid electric potential $[V]$& \\ 
$\phi_e^\pm$  &  Electrolyte electric potential $[V]$  &\\
$i^\pm$   &  Current density $[A/m^2]$  &   \\
$j^\pm$  & Molar ion flux $[mol/m^2s]$ &   \\      
$i_0^\pm $ & Exchange current density $[A/m^2]$ & \\  
$\eta^\pm$ & Overpotential $[V]$ &\\
$c_{s,e}^\pm$  & Lithium concentration at particle surface $[mol/m^3]$&  \\
$I $ & Applied current $[A/m^2]$&  \\ 
$V $ & Terminal voltage $[V]$&  \\ 
$D_s^\pm,\ D_e$ & Diffusivity of solid, electrolyte phase $[m^2/s]$ & \\
$t_{\pm}^{0}$ & Transference numbers of the cation and
anion $[-]$& \\
$\varepsilon_s^\pm,\varepsilon_e$ & Volume fraction of solid, electrolyte phase $[-]$ & \\ 
$F$ & Faraday’s constant $[C/mol]$ &\\
$\sigma^\pm$  & Bulk conductivity $[1/Ωm]$ &  \\
$\kappa$ & Ionic conductivity $[1/Ωm]$ &  \\   
$R $ & Universal gas constant $[J/mol-K]$& \\ 
$T$ & Temperature $[K]$ &\\
$f_{\pm}$  & Mean molar activity coefficient $[-]$&   \\
$a_{s}$  & Specific interfacial surface area $[m^2/m^3]$ &   \\     
$\alpha$  &  Charge-transfer coefficient $[-]$ &\\
$\kappa_{0}^\pm$   &  effective reaction-rate constant\ $[(mol^{\alpha-1}m^{4-3\alpha} s^{-1})]$&   \\
$c_{s,\max}^\pm $ & Maximum concentration of solid material $[mol/m^3]$& \\  
$U^\pm$  & Open circuit potential of solid material $[V]$&\\
$R_s^\pm$  & Particle radius in solid phase $[m]$ &   \\ 
$L$  & Length of region  &   \\ 
\cline{1-2}
\end{tabular}
\end{table}




\bibliography{bibliography}             








\end{document}